\documentclass[mathleft]{an}
\usepackage{graphicx}
\usepackage{times}
\usepackage[T1]{fontenc}

\begin{document}

\Pagespan{1}{}
\Yearpublication{}%
\Yearsubmission{}%
\Month{}%
\Volume{}%
\Issue{}%

\title{The Brightness of the Galactic Radio Loops
at 1420 MHz: \\
Some Indications for the Existence of Loops V and VI}

\author{V. Borka\inst{1}\thanks{Corresponding author:
  \email{vborka@vin.bg.ac.yu}\newline}
,\ J. Milogradov-Turin\inst{2}\thanks{
\email{jelenam@matf.bg.ac.yu}} \and D. Uro\v{s}evi\'c
\inst{2}\thanks{ \email{dejanu@matf.bg.ac.yu}}\fnmsep\thanks {This
research is part of the project "Gaseous and stellar component of
galaxies: interaction and evolution" (No. 146012) supported by the
Ministry of Science and Environmental Protection of Serbia.}}

\titlerunning{Brightness of Galactic Radio Loops
at 1420 MHz: Some Indications for Loops V and VI Existence}
\authorrunning{Borka et al.}

\institute{Laboratory of Physics (010), Vin\v ca Institute of
Nuclear Sciences, P. O. Box 522, 11001 Belgrade, Serbia \and
Department of Astronomy, Faculty of Mathematics, University of
Belgrade, Studentski trg 16, 11000 Belgrade, Serbia}

\received{2006 September 10}
\accepted{2008 January 15}
\publonline{}


\abstract{In this article we use 1420 MHz data to demonstrate the
likely reality of Galactic radio Loops V and VI. We further estimate
distances and spectral indices for both these and the four main
radio loops. In the cases of Loops I - IV, radio spectral indices
are calculated from the mean brightnesses at 1420 and 820/404 MHz.
The spectral indices of Loops V and VI are obtained from $T - T$
plots between 1420 and 408 MHz. Using the supernova remnant (SNR)
hypothesis for the origin of radio loops, distances are calculated
from the surface brightnesses and the angular diameters at 1420 MHz.
We also study how results for brightnesses and distances of radio
loops agree with current theories of SNR evolution. For this
purpose, the ambient density and initial explosion energy of the
loops are discussed. We also discuss applications of different
$\Sigma - D$ relations. The results obtained confirm a non-thermal
origin and nearby locations for the Galactic radio loops. Therefore,
we have indications that they are very old SNRs that evolve in low
ambient densities, with high initial explosion energies.}

\maketitle

\section{Introduction}

It is well known that some spurs seen in the radio continuum
background distribution can be associated such as to trace small
circles. The set of spurs which form a particular small circle is
called a loop. Four major loops had been recognized by the early
seventies, these being discovered and studied in the following
order: Loop I (Large, Quigley \& Haslam 1962; Haslam, Large \&
Quigley 1964; Large, Quigley \& Haslam 1966; Salter 1970), Loop II
(Large et al. 1962; Quigley \& Haslam 1965; Salter 1970), Loop III
(Quigley \& Haslam 1965, Salter 1970) and Loop IV (Large et al.
1966; Salter 1970; Reich \& Steffen 1981). The most precise
determination of parameters for these circles was made by Salter
(1970) and published by Berkhuijsen, Has\-lam \& Salter (1971).
Salter used the then best data available at 408, 404, 240 and 178
MHz. A detailed review of the subject was published by Salter
(1983). It was noticed by Milogradov-Turin (1970, 1972, 1982) and
Milogradov-Turin \& Uro\v{s}evi\'c (1997) that some other spurs
could be connected into loops. They proposed Loop V, formed by
negative latitude spurs in Taurus, Pisces and Pegasus, and Loop VI,
corresponding to weak positive latitude spurs in Leo and Cancer.
Loop V has well defined "end" spurs and weak "bridges" between them,
while Loop VI is practically equivalent to the high brightness
temperature region betwe\-en cold holes near the North Galactic
Pole. They further computed parameters for the proposed loops using
the survey at 38 MHz by Milogradov-Turin \& Smith (1973).

\begin{figure}
\centering
\includegraphics[width=0.48\textwidth]{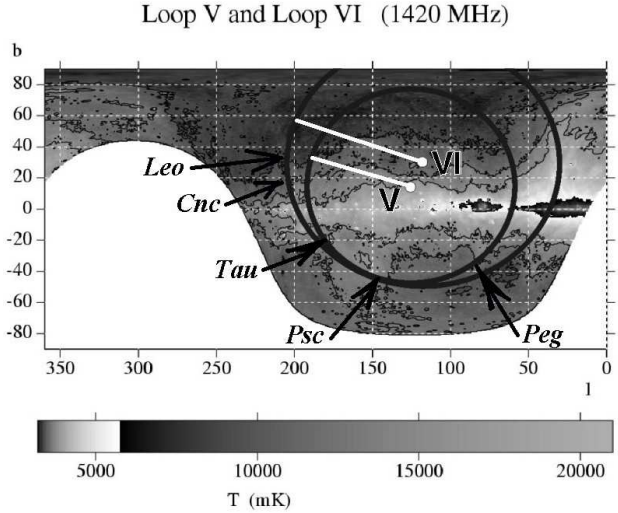}
\caption{The 1420 MHz northern-sky continuum survey. The contour
lines represent brightness temperatures of 3400, 3600 and 3800 mK.
The spurs in Taurus, Pisces and Pegasus form Loop V and the spurs in
Leo and Cancer form Loop VI. The loop centers are marked by small
white circles. The temperature scale in mK is given below the
figure.}
\label{fig01}
\end{figure}

The radio surveys at 1420 MHz (Reich \& Reich 1986) and at 408 MHz
(Haslam et al. 1982) are used as the basic source of data in the
present paper. The northern-sky map of the radio continuum
brightness at 1420 MHz (data are taken from Reich \& Reich 1986) is
presented in Figure \ref{fig01}. The paths of Loops V and VI are
marked in order to show the approximate positions and dimensions of
these loops. The small circles represent data from Table 1 of
Milogradov-Turin \& Uro\v{s}evi\' c (1997) which contains
approximate parameters for the best-fit circles. Although other
theories ha\-ve been suggested for the origins of the radio loops,
(for example, Sofue 1994, 2000 suggested that the radio loops and
the diffuse X-ray emission around the Galactic center can be
explained by a giant explosion in the Galactic center region), we
present results which suggest that radio Loops V and VI may
represent supernova remnants (SNRs).

Recent $\Sigma - D$ (surface brightness -- diameter) relations
(Uro\v{s}evi\'c 2002; Arbutina et al. 2004) have been used in this
paper to estimate distances for the Galactic radio loops.

\section{The Data}

The survey at 1420 MHz (Reich \& Reich 1986) provided a basis for
the recomputing of small circles for the major loops and Loops V and
VI. The observational data were obtained with the Stockert 25 m
telescope. The effective sensitivity is about 50 mK in
T$_\mathrm{b}$ (where T$_\mathrm{b}$ is for an ave\-rage brightness
temperature), and the angular resolution is 35$^\prime$. The survey
at 408 MHz (Haslam et al. 1982) is an all-sky survey obtained by
combining surveys from four telescopes, the Jodrell Bank Marks 1 and
1A, Effelsberg and Parkes, with effective sensitivity of about 1.0 K
in T$_\mathrm{b}$ and an angular resolution of 51$^\prime$. These
data are available from the MPIfR Survey Sampler
("Max-Planck-Institut f\"{u}r Radio\-astro\-nomie", Bonn), which is
available on-line at: http:// www.mpifr-bonn.mpg.de/survey.html~ and
allows users to pick a region of the sky and obtain images at a
number of wavelengths.

\section{Analysis}

One way of understanding SNR evolution is by studying the relation
between the surface brightnesses and diameters of SNRs, the
so-called $\Sigma - D$ relation. Shklovskii (1960) developed the
first theoretical $\Sigma - D$ relation, which is assumed to be of
the form:

\begin{equation}
\Sigma = A D^{- \gamma},
\end{equation}

\noindent where $D$ is the diameter of the object, and $A$ is a
coefficient which does not depend on $D$, but does depend on the
initial energy that the supernova provided to the remnant and on the
density of unperturbed matter through which the remnant is
expanding. $\gamma $ is a function of $\alpha$, the flux density
spectral index $(S_\nu\propto\nu^{-\alpha})$.

The first empirical $\Sigma  - D$ relation was derived by Milne
(1970), while Berkhuijsen (1973) was the first person to include the
Origem Loop and Loop I in the $\Sigma - D$ relation. All the main
radio loops (Loops I, II, III and IV) were included in the study of
the $\Sigma - D$ relation of SNR at known distances by Berkhuijsen
(1986).

Brightness temperatures over the areas of the loops, as required for
our $T - T$ plots, were obtained from the continuum survey data of
Reich \& Reich (1986; at 1420 MHz) and Haslam et al. (1982; at 408
MHz). Each loop area was divided into different sections
(corresponding to spurs) and estimates for these sections were
combined. The areas used for the individual loops were not defined
with respect to their best-fit circles, but were obtained from the
radio continuum maps. Assuming that Loops I-VI are old SNRs,
pro\-bably in their isothermal (radiative) phase, only the data from
the relevant spurs of the loops were used for deriving brightness
temperatures. The surface brightnesses of the SNRs must be above the
sensitivity limit of the observations and must be clearly
distinguishable from the Galactic background emission (Green 1991).
Therefore, the data from the fainter parts of the loop (which have
very low surface brightness) are not taken into account because it
is very difficult to resolve them from the background. On the other
hand, this would significantly reduce the brightnesses of entire
loops and there is a general trend that fainter SNRs tend to be
larger (Green 2005). Consequently, the $\Sigma - D$ relations that
were applied could give overestimated values for the diameters. For
the estimation of brightness temperatures over the spurs we had to
take into account background radiation (Webster 1974). Borders
enclosing the spurs are defined such as to clearly delineate a spur
and its background. For the method of calculation see Borka (2007).

The loops are believed to be SNRs whose material is expanding inside
of bubbles of low density, created either by earlier SNR explosions
or by strong stellar winds (Salter 1983; McKee \& Ostriker 1977 and
references therein).

The locations of the Galactic radio loops (and their spu\-rs) in the
sky, and the values of their average brightness temperatures and
surface brightnesses, can be found in Borka (2007; Tables 1 and 2
therein).

The temperatures at 1 GHz were calculated using the temperatures at
1420 MHz given by Borka (2007) and the spectral indices of
Berkhuijsen (1973) for Loops I-IV, plus those calculated here for
Loops V and VI. We calculated the surface brightness by applying the
following relations:

\begin{equation}
T_{1\mathrm{GHz}} / T_{1.42\mathrm{GHz}} = (1.0 / 1.42)^{- \beta}
\end{equation}

\begin{equation}
\Sigma_{1\mathrm{GHz}} = 2k T_{1\mathrm{GHz}} (1\mathrm{GHz} /
c)^{2}
\end{equation}
\noindent where $\beta $ is the temperature spectral index.

In order to determine distances for the Galactic radio loops I-VI,
we applied the surface brightnesses from Borka (2007) to the recent
$\Sigma - D$ relations of Uro\v{s}evi\'c (2002) and Arbutina et al.
(2004). In addition, the values of the spectral indices of the main
Galactic Loops I-IV were obtained directly from spectra of
temperature versus frequency. To produce these spectra, with the
1420 MHz data from Borka (2007) (map of Reich \& Reich 1986) we used
data from Berkhuijsen (1973) at 820 MHz for Loops I-III, and at 404
MHz for Loop IV. The survey at 820 MHz is by Berkhuijsen (1972) and
that at 404 MHz by Seeger et al. (1965). The $T - T$ method was used
for the determination of spectral indices of Loops V and VI.

Diameters were calculated using the $\Sigma - D$ relations
(Uro\v{s}evi\' c 2002; Arbutina et al. 2004):

\begin{equation}
\Sigma_{1\mathrm{GHz}} = 8.43\cdot 10^{- 18} D^{- 2.06}
\label{equ04}
\end{equation}

\begin{equation}
\Sigma_{1\mathrm{GHz}} = 2.5\cdot 10^{- 15} D^{- 3.4}
\label{equ05}
\end{equation}

\noindent ($\Sigma_{1\mathrm{GHz}}$ in
W\,m$^{-2}$\,Hz$^{-1}$\,Sr$^{-1}$ and $D$ in pc) and from these the
distances to the centers of the shells were calculated using the
relation:

\begin{equation}
r = D / (2 \sin \theta)
\end{equation}

\noindent with angular radii ($\theta )$ taken from Milogradov-Turin
\& Uro\-\v{s}evi\' c (1997). In Uro\v{s}evi\'c (2002) the Galactic
loops are ad\-ded to the set of 157 calibrators, and relation
(\ref{equ04}) was derived. In Arbutina et al. (2004) a good $\Sigma
- D$ relation (\ref{equ05}) was derived for the M82 starburst galaxy
and they suggest that a similar relation might exist for Galactic
SNRs associated with giant molecular clouds.

In addition, we have the relation:

\begin{equation}
T_{\mathrm{b}} = K \nu ^{-\beta}
\end{equation}
\noindent where $\beta $ is the spectral index and $K$ is a
constant. Knowing the values of brightness temperature at two
frequencies, we were able to calculate spectral indices.

For Loops V and VI, spectral indices were calculated using $T - T$
plots. These were made between temperatures at 1420 MHz (Reich \&
Reich 1986) and 408 MHz (Haslam et al. 1982).

By the relation:

\begin{equation}
dT_{\mathrm{b}}(\nu_{1}) / dT_{\mathrm{b}}(\nu_{2}) = (\nu_{1} /
\nu_{2})^{- \beta}
\end{equation}
\noindent we can determine the temperature spectral index, $\beta $,
from the slope of straight-line fitted to a $T - T$ plot. Using the
above relations, diameters and distances were calculated for Loops V
and VI.

\begin{table}
\centering
\caption{The diameters $D$ (pc) and distances $r$ (pc) to
the main radio loops I-IV derived from the $\Sigma - D$ relations
given by Uro\v{s}evi\'c (2002) and Arbutina et al. (2004),
respectively.}
\begin{tabular}{|l|c|c|c|c|}
\hline
\raisebox{-1.50ex}[0cm][0cm]{Loop} &
\multicolumn{2}{|c|}{derived from (\ref{equ04})} & \multicolumn{2}{|c|}{derived from (\ref{equ05})} \\
\cline{2-5} & $D$ (pc) & $r$ (pc) & $D$ (pc) & $r$ (pc) \\
\hline \hline
\textbf{ I} & 169 $\pm $ 38 & 99 $\pm $ 25 & 119 $\pm $ 17 & 70 $\pm $ 12 \\
\hline
\textbf{ II} & 187 $\pm $ 42 & 129 $\pm $ 33 & 127 $\pm $ 18 & 88 $\pm $ 16 \\
\hline
\textbf{ III} & 167 $\pm $ 37 & 150 $\pm $ 42 & 118 $\pm $ 17 & 106 $\pm $ 21 \\
\hline
\textbf{ IV} & 310 $\pm $ 69 & 462 $\pm $ 103 & 172 $\pm $ 24 & 257 $\pm $ 50 \\
\hline
\end{tabular}
\label{tab01}
\end{table}

\begin{table}
\centering
\caption{The spectral indices of the main radio loops.}
\begin{tabular}{|l|c|}
\hline Loop & $\beta$ \\
\hline \hline
\textbf{ I} & 2.69 $\pm $ 0.05 \\
\hline
\textbf{ II} & 2.91 $\pm $ 0.05 \\
\hline
\textbf{ III} & 2.59 $\pm $ 0.04 \\
\hline
\textbf{ IV} & 2.65 $\pm $ 0.09 \\
\hline
\end{tabular}
\label{tab02}
\end{table}

\begin{table}
\centering
\caption{The diameters $D$ (pc) and distances $r$ (pc) to
the radio loops V and VI derived from the $\Sigma - D$ relations
given by Uro\v{s}evi\'c (2002) and Arbutina et al. (2004),
respectively.}
\begin{tabular}{|l|c|c|c|c|}
\hline
\raisebox{-1.50ex}[0cm][0cm]{Loop} &
\multicolumn{2}{|c|}{derived from (\ref{equ04})} & \multicolumn{2}{|c|}{derived from (\ref{equ05})} \\
\cline{2-5} & $D$ (pc) & $r$ (pc) & $D$ (pc) & $r$ (pc) \\
\hline \hline
\textbf{ V} & 235 $\pm $ 36 & 124 $\pm $ 22 & 146 $\pm $ 22 & 77 $\pm $ 14 \\
\hline
\textbf{ VI} & 255 $\pm $ 39 & 134 $\pm $ 23 & 153 $\pm $ 23 & 80 $\pm $ 14 \\
\hline
\end{tabular}
\label{tab03}
\end{table}

\begin{table}
\centering
\caption{The spectral indices of radio loops V and VI,
between 1420 and 408 MHz.}
\begin{tabular}{|l|l|c|c|}
\hline
Loop& ~~~Spur& $\beta$ & $ \langle \beta \rangle $ \\
\hline \hline
\raisebox{-3.00ex}[0cm][0cm]{\textbf{ V}}& Taurus& 2.89 $\pm $ 0.10& \raisebox{-3.00ex}[0cm][0cm]{$2.90 \pm 0.11$} \\
\cline{2-3} & Pisces& 2.63 $\pm $ 0.12& \\
\cline{2-3} & Pegasus& 3.12 $\pm $ 0.09& \\
\hline \raisebox{-1.50ex}[0cm][0cm]{\textbf{ VI}}& Leo& 2.91 $\pm $ 0.13& \raisebox{-1.50ex}[0cm][0cm]{$2.77 \pm 0.13$} \\
\cline{2-3} & Cancer& 2.71 $\pm $ 0.12& \\
\hline
\end{tabular}
\label{tab04}
\end{table}

\section{Results}

The results for the four main loops are presented in Tables
\ref{tab01} and \ref{tab02}. The diameters and distances to the
radio loops derived from the $\Sigma - D$ relations given by
Uro\v{s}evi\'c (2002) and Arbutina et al. (2004) are shown in Table
\ref{tab01}. The spectral indices for the four main radio loops,
calculated as described above, are shown in Table \ref{tab02}.
Spectra were generated using mean temperatures at two different
frequencies as needed for making the simplest linear fits. The two
frequencies are 820 and 1420 MHz for Loops I-III, and 404 and 1420
MHz for Loop IV. All earlier determinations of the radio loop
spectral indices, except for this paper and Borka (2007), were based
on $T - T$ methods. Besides, the values obtained (all greater than
2.2) confirm a non-thermal origin for the emission from the radio
loops. The values for the brightness temperature spectral indices of
the loops are rather steep (about $2.5 < \beta < 3.0 $). This is at
the high end of the spectral index distribution for SNRs (Clark \&
Caswell 1976).

\begin{figure}
\centering
\includegraphics[width=0.48\textwidth]{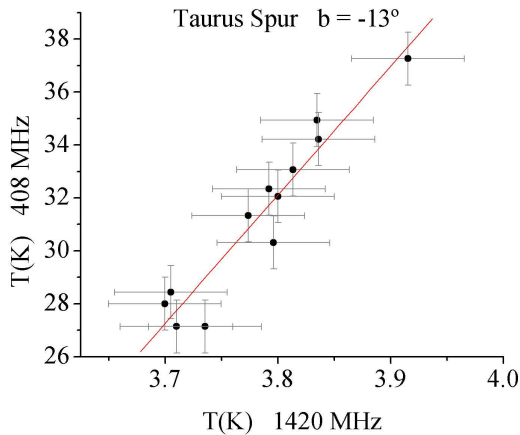}
\caption{The $T - T$ plot for the Taurus spur (Loop V) at $b =
-13^\circ$. $T$ (K) at 408 MHz is plotted versus $T$ (K) at 1420
MHz, for the longitude range from 178$^\circ$ to 189$^\circ$.}
\label{fig02}
\end{figure}

\begin{figure}
\centering
\includegraphics[width=0.48\textwidth]{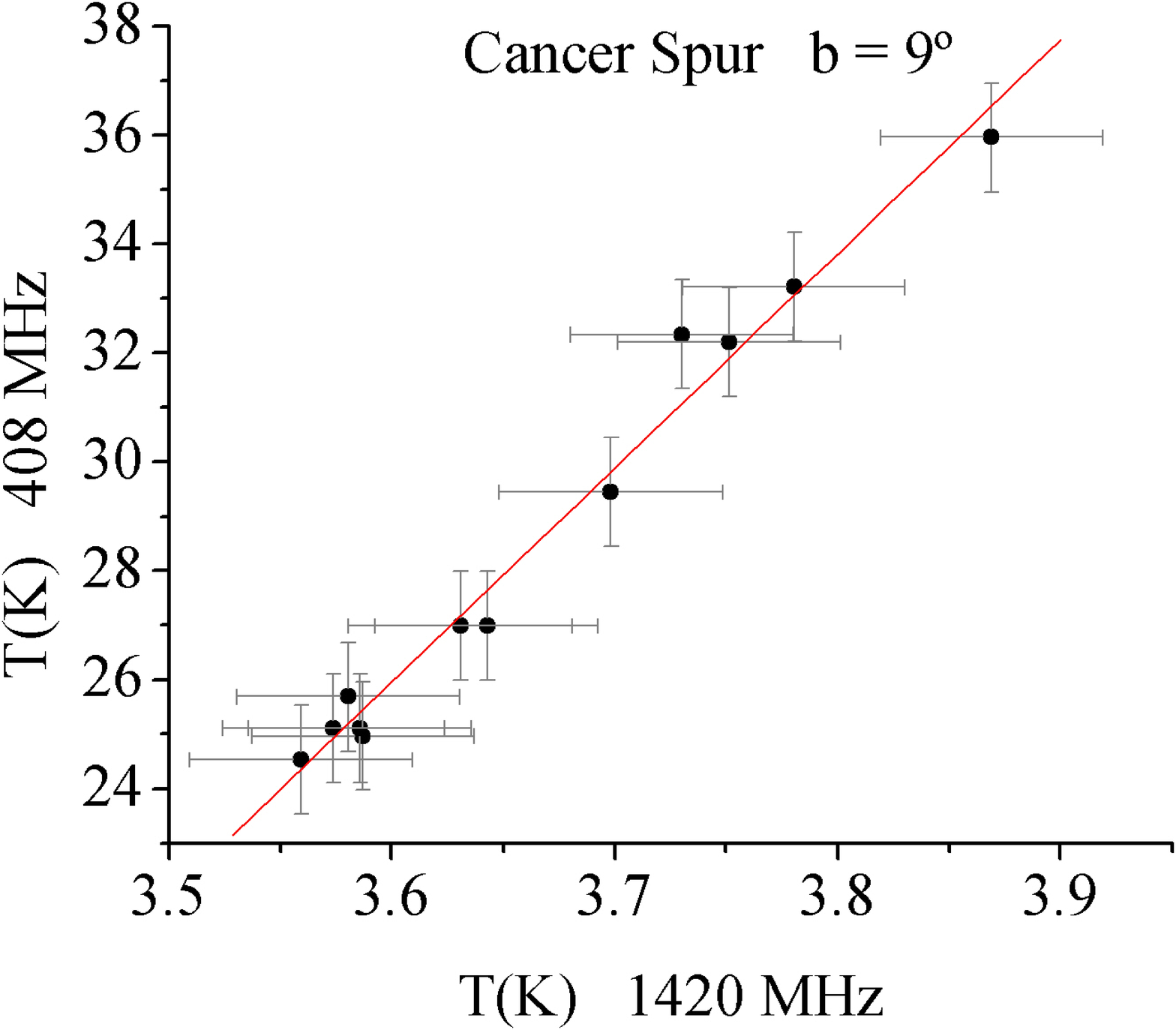}
\caption{The $T - T$ plot for the Cancer spur (Loop VI) at $b =
9^\circ$. $T$ (K) at 408 MHz is plotted versus $T$ (K) at 1420 MHz,
for the longitude range from 196$^\circ$ to 207$^\circ$.}
\label{fig03}
\end{figure}

The diameters of, and distances to, Loops V and VI derived using the
$\Sigma - D$ relations given by Uro\v{s}evi\'c (2002) and Arbutina
et al. (2004) are presented in Table \ref{tab03}. Representative $T
- T$ plots between 1420 and 408 MHz are given in Figures \ref{fig02}
and \ref{fig03}, with one plot for each loop. Using similar plots,
we determined spectral indices for all spurs which create Loops V
and VI. The spectral indices of Loops V and VI are shown in Table
\ref{tab04}. The values of spectral index derived from $T - T$ plots
for these two loops are obtained independently of the values for the
indices derived from the mean brightness temperatures in Borka
(2007).

We note that the values of the spectral indices in Tables
\ref{tab02} and \ref{tab04} are slightly different to those derived
by Borka (2007). However, the results agree within the absolute
error limits.

\section{Discussion}

We will assume below that all the radio loops discussed here are
SNRs (Berkhuijsen, Haslam \& Salter 1970; Berkhuijsen 1971;
Shklovskii \& Sheffer 1971; Salter 1983). The geometries of Loops V
and VI are presented in Figure \ref{fig01}. It is seen that Loop V
is projected entirely within Loop VI indicating a possible close
connection between these loops. Here, we will consider three
possibilities for such a connection:
\newline
(i) Milogradov-Turin \& Uro\v{s}evi\'c (1997) considered the
po\-ssibility that all the spurs of Loops V and VI define a unique
small circle corresponding to a single feature -- Loop V+VI.
However, their geometric fit did not provide reliable evidence for
this hypothesis.
\newline
(ii) a second possibility is that Loop V represents a "wrinkle" on
the shell of Loop VI, tracing part of a small circle on that shell.
If so, this small circle is oriented such that at one point (i.e. $l
= 150 ^\circ$, $b= - 44 ^\circ$) it meets the (different) small
circle representing the locus of tangential points from the Sun to
the Loop VI shell. Borka (2007) performed the first spectral
investigation of Loops V and VI using the mean brightness
temperatures at three frequencies (1420, 820 and 408 MHz) to form
the linear spectra for these loops. As shown in Table 1 of Borka
(2007), spurs of these loops belong to different and not overlapping
temperature intervals at all three frequencies. This suggests that
these spurs form two independent features, namely Loops V and VI.
\newline
(iii) a similar situation to that detailed in (i) and (ii) but
concerning Loops I and IV, was considered earlier by a number of
authors (Large et al. 1962; Davies 1964; Guidice 1971). However,
Borken and Iwan (1977) and Iwan (1980) proposed that the Loop I SNR
is very old, well into its radiative phase, and is being reheated by
the intersecting shock of Loop IV. They concluded that Loop IV has a
low density shell, is not yet in the radiative cooling phase, and
has not lost much energy before reheating the Loop I cavity. By
analogy, and given the similar distances and diameters derived for
these two loops (Table \ref{tab03}), we assume that the Loop V SNR
exploded within the previous (low density) cavity left by the Loop
VI SNR. In view of the similarity of the paths of the Loops V and
VI, one possible scenario is suggested by the recent paper by
Wolleben (2007). This author proposes that the emission of two
independent synchrotron-emitting shells make up the feature known as
Loop I (the North Polar Spur). While the presence of more than one
supernova remnant has been invoked previously to explain some of the
features of Loop I, (and especially its strong X-ray emission),
Wolleben's model invokes the presence of a second, very old remnant
to explain certain features he finds in the large-scale distribution
of polarized continuum intensity.

We have used recent $\Sigma-D$ relations to determine the distances
of the various loops. Our results demonstrate that the distances
from Arbutina et al. (2004) do not agree with those derived
independently by Berkhuijsen (1986). This is probably because
relation (\ref{equ05}) is not applicable to the Loops, as these are
thought to be situated in a low density environment. An analysis as
to how the $\Sigma-D$ relation depends on the density of the
interstellar medium is given in Arbutina \& Uro\v{s}evi\'c (2005).
Therefore, we conclude that the radio loops are evolving in a low
density environment. However, we emphasize once again our assumption
of the SNR explanation for the radio loops. The distances and
diameters obtained using the $\Sigma - D$ relation of Uro\v{s}evi\'
c (2002) show good agreement when compared with those in Borka
(2007), which were calculated from the $\Sigma-D$ relation derived
by Case \& Bhattacharya (1998). However, the relation of Uro\v
sevi\' c is preferred to that of Case \& Bhattacharya because it
adds the Galactic loops to the set of 157 calibrators. Also, the
relation of Case \& Bhattacharya shows better agreement with the
diameters of SNRs at known distances (Berkhuijsen 1986) than does
that of Arbutina et al. (2004).

\begin{figure}[ht!]
\centering
\includegraphics[width=0.45\textwidth]{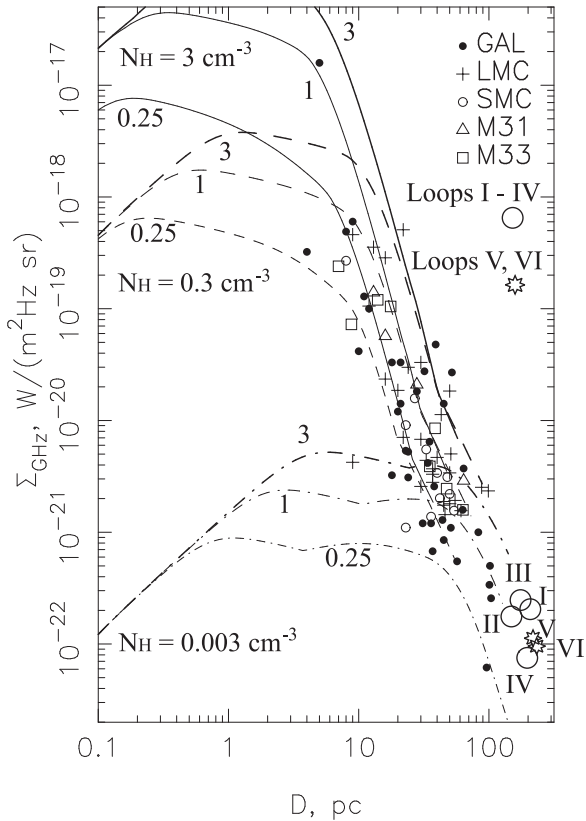}
\caption{The surface brightness to diameter diagram from Berezhko \&
V\"{o}lk (2004), with values for Loops I-VI added. Three different
densities for the ISM ($N_\mathrm{H}$ = 3, 0.3 and 0.003 cm$^{ {\rm
-} 3}$) are presented, plus three values for the SN explosion energy
($E_\mathrm{SN}$ = 0.25, 1 and 3 $\times 10^{ 51}$ erg).}
\label{fig04}
\end{figure}

Adopting diameters for the main loops from Berkhuijsen (1986) and
the diameters for Loops V and VI as estimated in this paper from
relation (\ref{equ04}), we can use the diagram of Berezhko \&
V\"{o}lk (2004) to determine the initial explosion energy that would
correspond to these loops, plus the density of the ambient
interstellar medium in their neighborhood. The $\Sigma - D$ diagram
at 1 GHz taken from Berezhko \& V\"{o}lk (2004) with the derived
values for the six loops superposed, is shown in Figure \ref{fig04}.
We identify Loops I-IV by circles and Loops V and VI by stars. For
Loops V and VI, this is a preliminary illustration. For additional
evidence of their SNR origin it would be necessary to calculate
distances (diameters) independently of the $\Sigma - D$ relation.
Then, from this $\Sigma - D$ diagram, it would be possible to
investigate the compatibility of the properties of Loops V and VI
with their being old SNRs. From Figure \ref{fig04}, it can be
concluded that the surrounding density of the radio loops is low and
that the explosion energy is high. Further, the locations of the
Loops on this $\Sigma - D$ diagram agree with the general
distribution and this strengthens the idea that they are old SNRs.

The main unsolved problems for the SNR hypothesis of the radio loops
are: (i) the high X-ray flux from the interior of Loop I, and (ii)
the high degree of spherical symmetry of all the loops. Both issues
are related to the evolutionary status of radio loops. If the loops
are SNRs, they are old and dimensionally immense (see Tables
\ref{tab01} and \ref{tab03}). High X-ray emission and spherical
symmetry are not expected from such SNRs.

The X-ray contours follow the North Polar Spur (NPS), but they are
located in the interior region of Loop I (Sofue 2000). We believe
that Loop I was in a late adiabatic phase, not so long ago, and in
that phase X-rays could be generated. At present, the NPS is
probably in an early radiative phase of expansion in which a thick
low temperature shell is generated (Chevalier 1974). The X-ray gas
is thermodynamically stable (the hot phase of the ISM) and can exist
for a million years, implying that X-rays can be observed from the
interior of Loop I. In a thicker uniform medium, a slow shock can
conserve its own approximately spherical form, but only for a very
short time interval. Due to the formation of a thick shell during
expansion in the radiative phase (Cox \& Reynolds 1987), the
velocity of the shock wave rapidly declines and the shell loses its
own identity. Generally, the radiative phase of evolution is short
in comparison to the adiabatic phase (e.g. McKee \& Ostriker 1977).

If a SNR is expanding in a hot bubble made by strong winds or
earlier supernova explosions, it has a chance to conserve an
approximately spherical form because of the uniformly distributed
matter formed by the evaporation of denser clouds (McKee \& Ostriker
1977). Hanayama \& To\-misaka (2006) concluded that a spherically
symmetric sha\-pe can be conserved if an old SNR evolves in an
environment with a weak magnetic field. In this case, the evolution
is similar to that of a non-magnetized SNR. In the pressure-driven
expansion stage, a SNR evolving in a medium with a magnetic field of
1 $\mu$G keeps its spherically symmetric shape (Hanayama \& Tomisaka
2006).

The above discussion suggests that the radio loops may ha\-ve a SNR
origin, but a definite answer needs much additional work.

\section{Summary}

The main results of this paper may be summarized as follows:
\newline
(i) We have estimated the distances of the main Galactic radio loops
and Loops V and VI from the northern-sky radio continuum survey at
1420 MHz (Reich \& Reich 1986) using two different $\Sigma - D$
relations.
\newline
(ii) The estimated distances of the main radio loops derived using
the 1420 MHz data are in good agreement with the earlier results
(e.g. Berkhuijsen 1986).
\newline
(iii) From the spectral index analysis we can confirm that the
emission from the radio loops is non-thermal in origin.
\newline
(iv) We have demonstrated the probable existence of Loops V and VI.
The distances and spectral indices of Loop V and Loop VI are
estimated. In general, they display shell structure and have
non-thermal spectra, similar to the main loops I-IV which are
usually assumed to be SNRs.

\acknowledgements

The authors are grateful to the referee for ve\-ry useful remarks
and suggestions which helped us improve the paper.

\end{document}